\documentclass[sigconf, auth]{acmart}

\begin{document}

\title{More Effective Software Repository Mining\\
{\footnotesize \textsuperscript{}}
}
\author{Adam Tutko}
\email{atutko@vols.utk.edu}
\orcid{0003-2412-5409}
\affiliation{%
  \institution{University of Tennessee}
  \city{Knoxville}
  \state{TN}
  \postcode{37996}
}
\author{Austin Z. Henley}
\email{azh@utk.edu}
\orcid{0003-1069-2795}
\affiliation{%
  \institution{University of Tennessee}
  \city{Knoxville}
  \state{TN}
  \postcode{37996}
}
\author{Audris Mockus}
\email{audris@utk.edu}
\orcid{0002-7987-7598}
\affiliation{%
  \institution{University of Tennessee}
  \city{Knoxville}
  \state{TN}
  \postcode{37996}
}

\begin{abstract} \label{0}
Background: Data mining and analyzing of public Git software repositories is a growing research field. The tools used for studies that investigate a single project or a group of projects have been refined, but it is not clear whether the results obtained on such ``convenience samples'' generalize. Aims: This paper aims to elucidate the difficulties faced by researchers who would like to ascertain the generalizability of their findings by introducing an interface that addresses the issues with obtaining representative samples. Results: To do that we explore how to exploit the World of Code system to make software repository sampling and analysis much more accessible. Specifically, we present a resource for Mining Software Repository researchers that is intended to simplify data sampling and retrieval workflow and, through that, increase the validity and completeness of data. Conclusions: This system has the potential to provide researchers a resource that greatly eases the difficulty of data retrieval and addresses many of the currently standing issues with data sampling.
\end{abstract}

\maketitle

\section{Introduction} \label{1}

When developing software, it is a common occurrence for developers to rely on the Git version control system\footnote{https://git-scm.com/}. This is done primarily to make collaboration between developers easier and because it provides a fail-safe against catastrophic errors. This reliance on Git means that the software development process generates a lot of publicly available data. This data could be immensely useful to Mining Software Repository (MSR) researchers in facilitating creation of tools to support developers. MSR researchers have realized this potential and the field of research has experienced rapid growth in recent years. However, due to the vastness of the field, there are a lot of difficulties faced when first retrieving data.

Due to the immensity of the Git ecosystem, one's ability to do any form of analysis often requires sampling of the data. However, if researchers are not careful when sampling, the data could easily be flawed. Due to the vastness of the dataset, it is easy to retrieve bad data or heterogeneous data \cite{Trautsch2016MSR}. This process is made even more difficult for MSR researchers focused on Git data because the retrieval process requires using difficult APIs or manual crawling of Git repositories. Many MSR papers currently use these methods to perform data retrieval and the difficulties with the retrieval process is well documented \cite{Kolovos2019MSR}, \cite{Sakamoto2012IWESE}.  


Alongside this, many MSR papers perform their sampling by manually selecting one or more projects. As seen in \cite{Yang2016ICMSR}, \cite{Silva2017MSR}, \cite{Yu2015MSR} the project selection can be based on a list of different criteria. \cite{Silva2017MSR} attempts to separate which projects to analyze by randomly selecting a project from a list of 20 randomly retrieved Java repositories hosted on GitHub that contain a Maven project file (\texttt{pom.xml}). On the other hand, \cite{Yang2016ICMSR} specifically selects five large scale repositories (OpenStack, LibreOffice, AOSP, QT, and Eclipse) that are integrated with Gerritt and Git. Similarly, there are many ways of performing the data extraction and often they do not follow a set process. As noted in \cite{Yang2016ICMSR}, the process can change even within one's own workflow since the process of scraping one repository is not always applicable to another. This lack of conformity in sampling and data extraction practices between researchers provides a potential for bias and errors in data retrieval. 



Many of the publicly available repositories are hosted on GitHub\footnote{https://github.com} (a popular code hosting and version control platform). As noted above, in order to select a sample of projects for analysis MSR researchers typically filter by the metadata available on GitHub. GitHub categorizes these projects based on metadata such as stars (a measure for users to keep track of repositories they like) or languages used. Currently, GitHub is the only well known system for sampling of projects and it is quite often the first choice of MSR researchers. For example, researchers might sample projects with more than three stars and/or further refine the search by using the language attribute, the project description, project creation data, or other metadata provided by GitHub API. Unfortunately, this process is time consuming, error prone (sometimes metadata is absent or incorrectly specified), and it is only possible to sample projects based on the very limited set of attributes provided by GitHub API. For example, there is no way to retrieve projects that contain a certain number of source code files from a specific language or to determine all projects that a specific developer has worked on. 

The World of Code (WoC)~\cite{Ma2019MSR} aims to provide an interface that simplifies the data retrieval process by allowing mass retrieval of git data that resides across all open source repositories such as commits, files, authors, projects, and blobs. WoC allows for rapid cross-data retrieval due to a set of key-value mappings between data types (e.g. commit to author of the commit). More information on the initial implementation of this interface can be found in \cite{Ma2019MSR}.
Such capabilities are particularly suited to support representative sampling and, more generally, research on software ecosystems. In this paper we describe how we used WoC to support sampling of repositories based on specific criteria that are not available via, for example, GitHub API (e.g. number of commits, number of committers, lines of code, languages used, age of the repository, library dependencies). Specifically, we compiled databases to provide these sampling capabilities based on activities in git repositories or by authors making commits to all public git repositories.

The compiled datasets are stored in a MongoDB collection to make extraction and analysis easier. For MSR researchers interested in analyzing a single project, WoC can be used for data retrieval beyond the scope of the project. Due to the key-value mappings between Git data types, it is possible to easily retrieve information about developer activities outside of the project being analyzed.

\section{Creation of the Databases for Sampling} \label{3}
Using the World of Code, a set of publicly available datasets was created to allow for more exact sampling of git project/author data. To make data retrieval less challenging for users not accustomed to the WoC system, the datasets are stored in MongoDB collections labeled as metadata. While these datasets contain useful information, they are not an exhaustive collection of all the possible data that can be derived from Git. The World of Code is a system that stores Git data in an easily accessed form. Alongside this, the system is not restricted solely to one source code hosting site. Instead, it saves data from the greater git ecosystem. Both of these metadata datasets were produced by iterating over the entirety of the projects/authors contained within the World of Code. Utilizing the base-mappings between datatypes, it was possible to retrieve and store all the desired data for these collections. However, considering the size of data contained with WoC, the process of creating these datasets takes about 24 hours for authors and roughly 48 hours for projects.

\begin{figure}[t]
\centering
\includegraphics[width=0.5\textwidth]{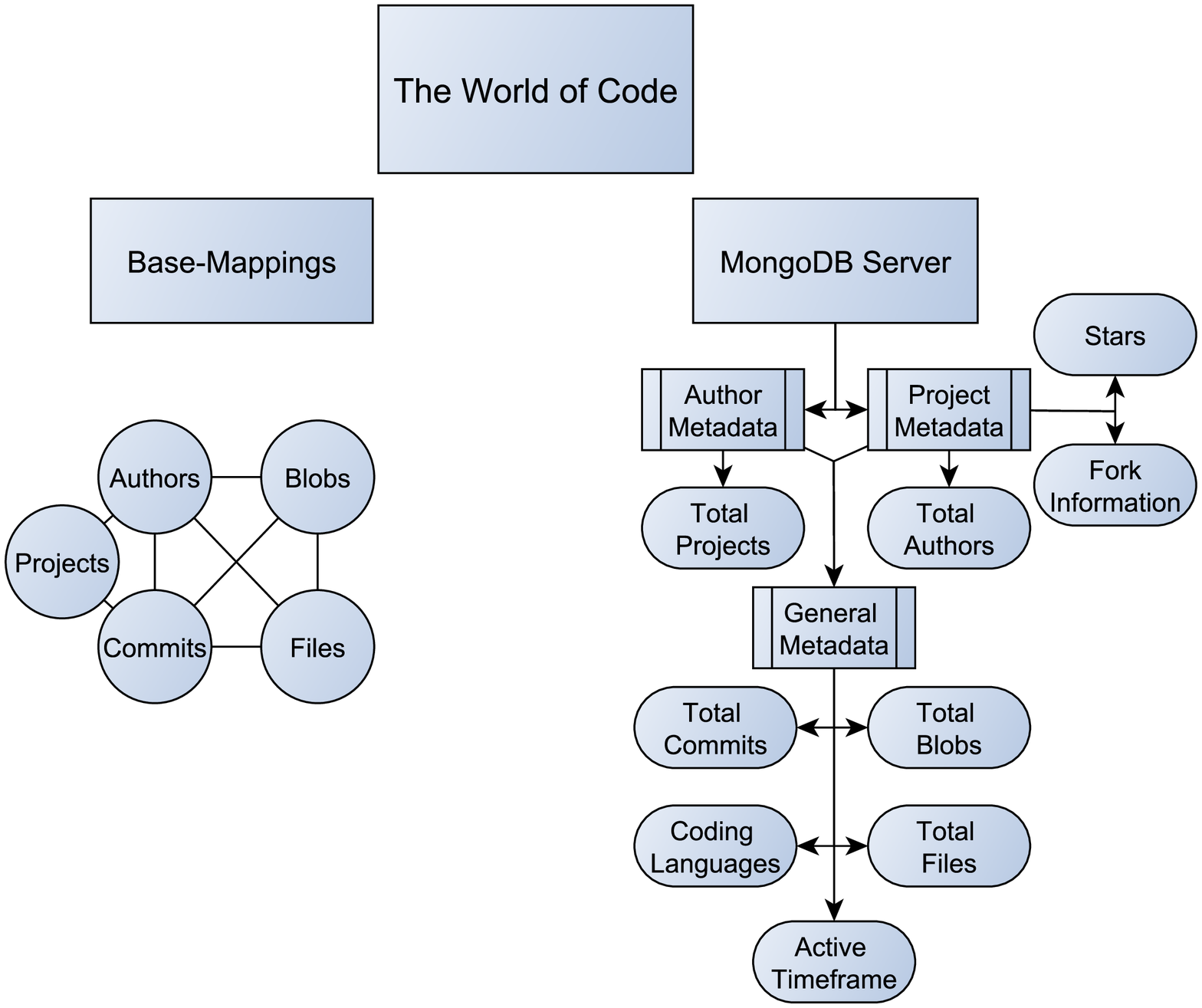}
\caption{System Layout for resources within the World of Code}
\end{figure}

\subsection{The types of MSR analyses that could benefit from better sampling}
To determine if the World of Code provides a system unique in its ability to be used for certain MSR research, we considered a set of usage examples. Furthermore, we recruited 3 external researchers to perform/provide research tasks. 

In the development process, a potentially large factor for developers when considering how to market their software is coding language popularity ( e.g. which language should the product be implemented in, which language should be provided for others to interface with a component, etc). One way to determine popularity using the World of Code is by plotting language usage over time. By assessing the file extensions of each file it is possible to determine which files are related to which coding languages. Afterwards, using the mapping from files to commits, the timestamp related to each commit can be used to determine usage per year. 

Another avenue of research the World of Code offers is analyzing developer ecosystems. When analyzing the developers in git, it is quite troublesome to retrieve accurate and specific data of developers names. Names or emails of developers are frequently misspelled, incomplete, or completely missing. This makes performing any research on developer networks quite difficult. However, the World of Code provides some options for enhancing accuracy measures when attempting to disambiguate author identities. To perform any form of disambiguation algorithm it is necessary to determine common patterns of data irregularity. Fortunately, the World of Code contains a near complete collection of author ids (e.g. John Doe JD@domain.com) within Git. This makes the dataset much more viable for studying and removing such irregularities. Furthermore, using WoC it is possible to even further disambiguate based on other factors of similarity. Such similarity measures include time patterns of commits, writing style within commits, and author id comparison within file changes.

Another form of research that WoC can help perform is analysis of sustainability of open source ecosystems. By analyzing the existing projects within a particular ecosystem, for example Python's PyPI packages, it is possible to determine how often these projects achieve "feature completeness" (achieved intended usage and requires no further maintenance) versus how often projects are abandoned. This can be done once again using the file to commit mapping and by assessing the imports within each file.

Other research areas include analysis of developers across project ecosystems, file cloning across ecosystems, repository filtering, the popularity and relationship between NPM packages, and investigating what influences adopting of software.

\subsection{A Use Case for project/author sampling}

\begin{figure}[h]
\centering
\includegraphics[width=0.3\textwidth]{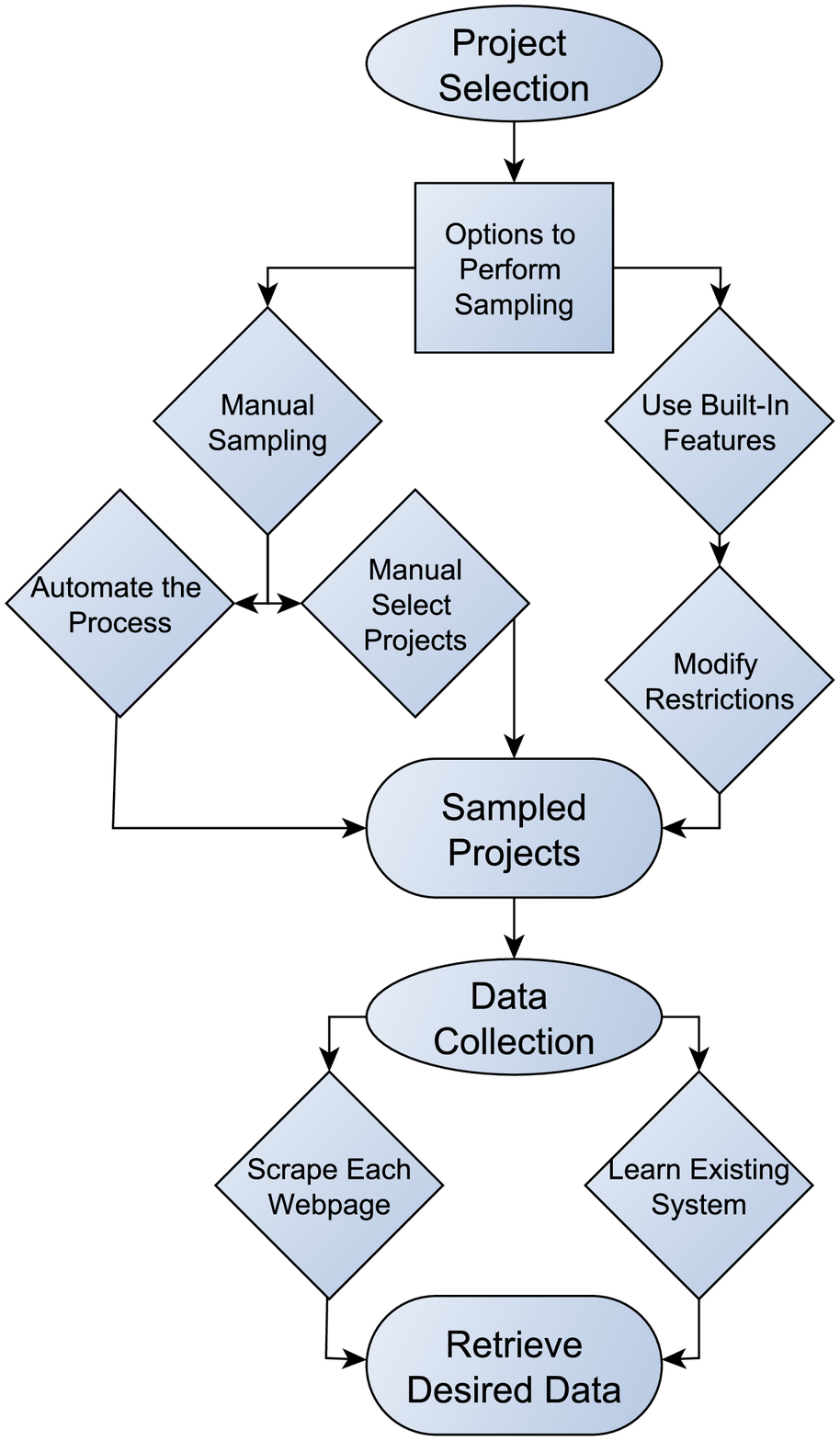}
\caption{Sampling/Data Retrieval Workflow without the World of Code}
\end{figure}

\begin{figure}[h]
\centering
\includegraphics[width=0.1\textwidth]{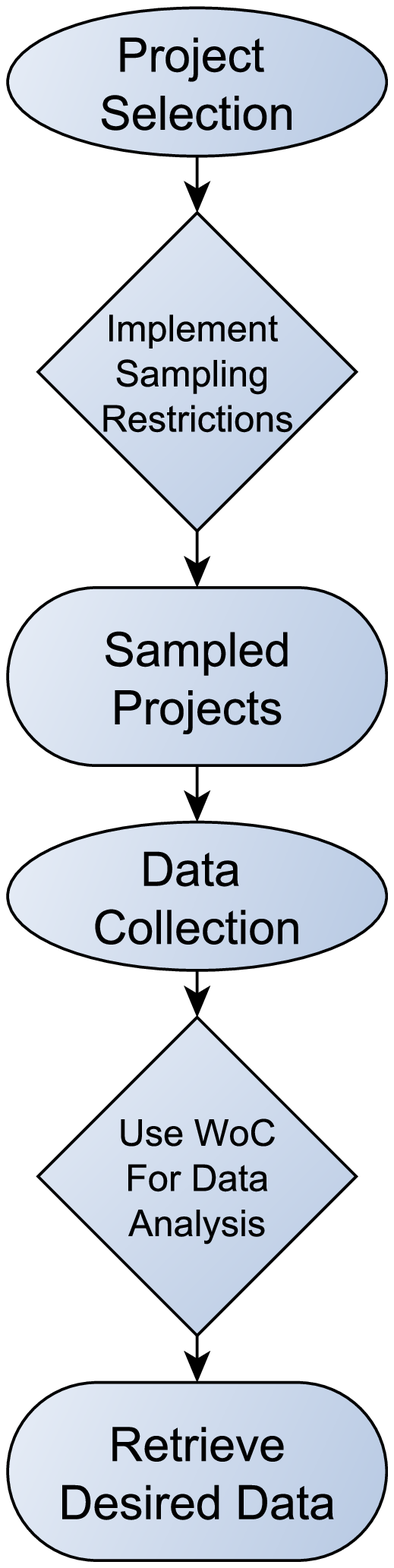}
\caption{Sampling/Data Retrieval Workflow without the World of Code}
\end{figure}

To further elucidate how Git data retrieval can be tedious, consider a hypothetical researcher Sam who wants to start repository analysis. Sam wants to analyze the frequency of changes to coding language files in projects. Considering that the average number of lines of code in C files is much greater than in Python files, Sam wants to determine whether C files are changed more often than Python files. To proceed with this research Sam needs to have access to all changes made to previous files over a timeframe in order to be able to determine which (if any) coding language experiences more frequent changes. Sam's sampling and data retrieval workflow is illustrated in Figure 2.

To achieve this goal, Sam wants to sample 20 disparate projects that contain at least 20 Python files each and a minimum of 5 committing authors. Sam also wants 20 projects that contains at least 20 C files and at least 5 committing authors. They also want these files to contain at least 50 lines of code for the Python files and over 100 lines for the C files. Sam has used GitHub in the past and believes that it must host many such projects. Thus, they want to use it when performing the project sampling. They do a quick search and finds that they can get a list of all currently trending C/Python projects. However, despite having a list of projects, they have no way to know which of these projects satisfy all of the requirements without directly assessing each project. 

Sam hopes to avoid such a time consuming process and wants to automate some of the work. Thus, they do another quick search and realizes that GitHub has a way to further restrict the resulting repositories based on the project metadata GitHub collects. Unfortunately, they quickly realize that some of the restrictions they need have not been implemented. Sam has no way to specifically restrict the results by the number of coding language files within each project. Furthermore, there does not seem to be a way to restrict the files to a number of lines. 

Thus, they need to decide how else they would like to restrict the search or choose to manually search through each project. To make their work more expedient, they decide to modify the search field slightly by simply choosing projects that have a large number of stars and fall into either the C or Python category. This will hopefully satisfy the requirements regarding the number of files and committing authors. Unfortunately, the workload of manually looking for 40 projects is still too time consuming. Thus, Sam decides to use the first 10 of each language found. 

After determining the projects Sam would like to analyze, they have to determine how to go about retrieving the desired data. They can either decide to manually scrape the data from each project's webpage or look for a system in place that will do it for them. They do a quick search and find that there are publicly available API that will allow them to interface with the GitHub project data. However, all of the API will take time to learn and it is hard to determine if the system will have all the desired data.

Thus, Sam must decide whether to learn how to use one of these systems or implement a temporary system to personally scrape the data. To avoid potentially retrieving bad data by performing manual scraping, they decide to learn one of the available systems and attempt to retrieve the information. Now Sam must determine how to go about assessing the rate of change of each file. They know that they can look at different commits in Git and determine if there were changes made. Thus, they choose an API that allows them to retrieve that information and after a little time and effort they finally retrieve the desired data.

Had Sam known about the World of Code, they could have saved much time and effort. Sam's hypothetical workflow using the World of Code is illustrated in Figure 3. Using the World of Code they would be able to perform their project sampling in the exact way they originally desired and could have performed their analysis/data retrieval while performing the sampling. Since the World of Code has access to the files, committing authors, and lines of code in each project, they would have been able to perform all of their work quickly and within one system. Since the World of Code contains basemappings between data-types, performing the sampling is a simple matter. 

This hypothetical process could have been solved in its entirety using the World of Code. Had Sam used the World of Code, Sam would have needed to retrieve all the files associated with a project. Then, to perform a count of the specific language files contained within the project, they could have simply assessed the file extensions of each file. Once a project with the desired number of files had been found, they could have retrieved the number of committing authors of that project by utilizing the direct mapping between projects and authors. This basemapping would return a list of authors and whether there were 5 or more authors in the project would be easily discerned. Finally, they could determine whether the coding files contain the desired number of lines by using the mapping between projects and blobs. Blobs in WoC contain the actual contents of each file. Thus, counting the lines of the blobs can discern if the project satisfies the set requirements. Furthermore, once a project had been found that satisfies the requirements, it is possible to directly perform the analysis. Since the World of Code contains all the commits associated with each project and every blob is linked to a commit, it is possible to determine how often one file has been changed. New blobs are not created unless there were changes made to the file. Thus, counting the blobs of each of the language files will quickly retrieve the desired information, and would have solved Sam's problem. Not only is the workflow simpler to figure out when using the World of Code, but it also only requires the user understand one system.

Further, had Sam desired to research individual committing authors within Git, the workflow without the World of Code would have been quite a bit more difficult. Currently, there is no known system that allows for sampling of authors and thus the research would have required personal implementation for any automation of the process. To make matters even more difficult, many of the publicly available API lack the requisite data regarding authors to perform in depth research. This would have made their workflow quite a bit more difficult and almost certainly would have forced them to manually implement any scraping of data. However, because the World of Code contains basemappings for authors similar to projects, performing a sampling of specific authors based on certain restrictions is still easily performed. Additionally, the analysis can also be done within WoC when researching authors.

Considering much of these examples are focused around making researchers' workflow easier, it is worth noting that much of the above stated work is possible using the metadata datasets compiled using the World of Code as well. While it is not possible to determine lines of code using the metadata datasets, performing the sampling based on number of language files and authors within the projects is easily accomplished using MongoDBs built-in restrictions. This would allow researchers to fully avoid having a firm grasp on Perl, Python, or the Unix Command Line which is necessary to perform the sampling using the World of Code. Then the later data retrieval/analysis process could be performed using their preferred API.

\subsection{Project Metadata Extracted from Code Commits} 
The project dataset is stored in a MongoDB collection titled\\ \texttt{proj\_metadata} followed by the current version of WoC \\(e.g. \texttt{proj\_metadata.Q}). The collection stores the total number of authors, commits, and files associated with the project. It also stores an activity range for each project based on the Unix timestamp linked to the first and last commit. Alongside this, the data includes coding language usage based on the file extensions of each file in the project. Due to the prevalence of forked projects in Git, if WoC determines a project to be a fork then the original location of the forked project is included. When the project is hosted on GitHub and has a stars rating, the collection includes information on the stars rating of the project.
\subsection{Author Metadata}
Like the project dataset, the author dataset is stored in a MongoDB collection titled \texttt{auth\_metadata} followed by the current version of WoC (e.g \texttt{auth\_metadata.Q}). It includes the total number of commits, blobs, files, and projects the author has participated in. It includes a time frame the author was active based on the Unix timestamps of the first and last commit. Lastly, it includes coding language usage based on file extension.

\section{Challenges and Limitations of the World of Code}

\emph{World of Code Versioning:} Since the World of Code contains so much information, mass-updates of information must be done in increments. Due to this fact, updating the version of WoC often happens months apart. This is necessary because updating the basemaps requires computationally intensive work that takes a non-trivial amount of time. This versioning of WoC means the data is subject to the latest WoC update. Thus, information on currently active repositories/authors is also restricted by this update system. Timeframe analysis based upon the first and last commit time is therefore not suitable when the data must be truly current. 

\emph{Language Inclusion:} There are restrictions to the languages included in the metadata datasets. The languages included are Ada, C/C++, COBOL, CSharp, Erlang, Fml, Fortran, Go, Java, Javascript, JL, Lisp, Lua, Perl, PHP, Python, R, Ruby, Rust, Scala, SQL, and Swift. When counting the files, languages were determined using the file extensions on the filenames. If the filename does not have one of the extensions for these languages it is not counted as a program file. Alongside this restriction, languages that do not require a specific file extension will be ignored by the algorithm and counted as a regular file. To analyze languages not included in this set, researchers may need to generate a personal dataset that includes the language, or request the language be included in the next iteration.

\emph{Forks:} There are limitations to the fork information provided in the metadata collections. This information is based on an arbitrary clustering method developed for determining forks. This was done partly using commits that are associated with many projects. Since there is a timestamp for each commit, it is theoretically possible to find the earliest commit in a project, then see what project that timestamp is associated with, and then claim the current project was forked from that project. However, if such a process is followed, the total number of "un-forked" projects becomes much smaller than is reasonable. This is because many projects have dependencies on widely used projects. Thus, the clustering method that was used had to make determinations on whether the project truly is a fork or not. 

\section{Related Work} \label{2}
Retrieving Git data can be a convoluted process and, as is discussed in \cite{Yang2016ICMSR}, the use of domain specific API can be difficult. \cite{Yang2016ICMSR} describes a framework meant to make mining of code review repositories on Gerrit easier. This framework retrieves code review information from publicly available repositories and stores it in a more easily accessed format. Desired information is compiled by scripts that query the Gerrit API and then parse the returned JSON object into a more easily used format. The parsed data is then stored in a relational database so that interested parties can access the data more easily. Similar to the data compiled by the World of Code, this data is intended for researchers interested in repository analysis. However, unlike WoC, their collected data is targeted at Code Review repositories and thus is not applicable to general repository analysis. 

A common practice to expand the research area is to analyze data pulled from general Git projects. Due to its dominant presence as a source code host, GitHub tends to be one of the primary targets for data retrieval and analysis. There is a publicly available API for GitHub that leverages the REST API to return JSON objects with the requested information. However, as was discussed in \cite{Gousios2013MSR}, using the GitHub API can be difficult, is restricted to specific fields of research, and may lead to biased results if used incorrectly. The API's limitations include restrictions on the total number of requests that can be made per hour and the difficulties of parsing the results. On top of the restrictions already in place from the API, GitHub also only provides a subset of the total Git ecosystem. Despite its prevalence, GitHub only makes up a fraction of the total number of Git repositories. Thus, the data collected there fails to include many projects. 

Also presented in \cite{Gousios2013MSR}, GHTorrent, a project meant to mirror the GitHub event timeline and store the raw JSON for retrieval, is another framework that is often used for repository analysis. The system has been picked up by many researchers because it removes the restriction on number of requests per hour and also attempts to provide the results in an already parsed form. This data is stored in a relational database and a set of MongoDB collections meant for querying. However, because it is based strictly off the GitHub ecosystem, it still has an issue with restricting the data to GitHub alone. It also introduces a set of new issues which are discussed in \cite{Gousios2013MSR}.

\section{Conclusions} \label{4}
In this paper, we presented the World of Code and the metadata datasets that were compiled within. Further, we outlined the limitations of WoC and the metadata in regards to data integrity. This system has potential to become an excellent sampling resource for researchers interested in MSR and for data analysis of Git projects and Git authors. The World of Code has potential for much easier Git data retrieval and could be used to drastically simplify the MSR research workflow. Furthermore, the system can be used to create similar metadata datasets in the future to make sampling of data for researchers much simpler.

\bibliographystyle{plain}
\bibliography{main.bib}

\vspace{12pt}

\end{document}